\documentstyle[12pt]{article}
\newcommand{\be}{\begin{equation}}
\newcommand{\ee}{\end{equation}}
\newcommand{\ba}{\begin{eqnarray}}
\newcommand{\ea}{\end{eqnarray}}
\newcommand{\ban}{\begin{eqnarray*}}
\newcommand{\ean}{\end{eqnarray*}}

\def\bbox{{\,\lower0.9pt\vbox{\hrule \hbox{\vrule height 0.2 cm
\hskip 0.2 cm
\vrule  height 0.2 cm}\hrule}\,}}

\def\bbox{{\,\lower0.9pt\vbox{\hrule \hbox{\vrule height 0.2 cm
\hskip 0.2 cm
\vrule  height 0.2 cm}\hrule}\,}}

\begin{document}
\setlength{\unitlength}{1mm}

\title{{\hfill {\small } } \vspace*{2cm} \\
Can One Understand Black Hole Entropy without Knowing Much
about Quantum Gravity?}
\author{
Dmitri V. Fursaev\footnote{e-mail: fursaev@thsun1.jinr.ru}
}
\maketitle
\noindent
\centerline{{\em Joint Institute for
Nuclear Research,}}
\centerline{\em Bogoliubov Laboratory of Theoretical Physics,}
\centerline{\em  141 980 Dubna, Russia}
\bigskip

\noindent

\begin{abstract}
It is a common belief now that
the explanation of the microscopic origin of the Bekenstein-Hawking
entropy of black holes should be available in
quantum gravity theory, whatever this theory will finally look like.
Calculations of the entropy of certain black holes in  string theory
do support this point of view.
In the last few years there also appeared a hope
that an understanding of  black hole entropy may
be possible even without
knowing the details of  quantum gravity.
The thermodynamics of  black holes is a low energy phenomenon, so only
a few general features of the fundamental theory may be really important.
The aim of this review is to describe some of the proposals
in this direction and the results obtained.
\end{abstract}

\vspace{.3cm}

\baselineskip=.6cm

\newpage

\section{Introduction}
\setcounter{equation}0

Black holes are specific solutions of the Einstein equations which
describe regions of a space-time where the gravitational field is
so strong that nothing, including light signals, can escape them.
The interior of a black hole is hidden from an external observer.
The boundary of the unobservable region is called the horizon.

A black hole can appear as a result of the gravitational collapse
of a star.
In this case it quickly reaches a stationary state characterized
by a
certain mass $M$ and an angular momentum $J$. If the collapsing matter
was not electrically neutral a black hole has an additional
parameter, an electric charge $Q$. These are the only
parameters a black hole in the Einstein-Maxwell
theory can have.
Its metric in the most general case is the Kerr-Newmann
metric. This statement is known as the "no-hair"
theorem\footnote{References on this subject as well as an introduction
in black hole physics can be found in \cite{FrNo:b}.}.
If $\Omega_H$ is the angular velocity of the black hole at the horizon,
$\Phi_H$ is the difference of the electric potential at the horizon
and at infinity, then by using purely classical equations one
arrives at the following variational formula \cite{BCH:73}
\begin{equation}\label{i.1}
\delta M=T_H \delta S^{BH}
 +\Omega_H\delta J+\Phi_H \delta Q,
\end{equation}
\begin{equation}\label{i.01}
S^{BH}={1 \over 4G}{\cal A},~~~T_H={\kappa \over 2\pi}~.
\end{equation}
Here ${\cal A}$ is the surface area of the horizon and
$G$ is the Newton gravitational constant
\footnote{Here and in what follows we use
the system of units where $\hbar=c=k_B=1$ ($k_B$
is the Boltzmann constant), and follow the notations
adopted in \cite{MTW:73}. In particular
the Lorentzian signature is $(-,+,+,+)$.}.
The constant $\kappa$ is called the surface gravity. It characterizes
the strength of the gravitational field near the horizon.
Relation (\ref{i.1}) has the form of the
first law of thermodynamic where
$S^{BH}$ has the meaning of an entropy,
$T_H$ is a temperature, and $M$ is an internal energy.
The quantity $S^{BH}$
was introduced in \cite{Bekenstein:72}-\cite{Hawking:75}
and is called the Bekenstein-Hawking entropy.
Strictly speaking (\ref{i.1}) defines the entropy and the temperature
up to a multiplier. This multiplier is fixed from another considerations:
$T_H$ is defined as the temperature of the Hawking radiation
from a black hole \cite{Hawking:75}\footnote{The mechanisms which give rise to
the Hawking radiation or quantum evaporation of black hole
are analyzed in \cite{BMP}.}.

One can also find  an analogy with other
laws of thermodynamics. For instance, by considering classical processes
with black holes one can conclude that the area of the horizon
never decreases, the observation which is reminiscent to the
second law. In quantum theory this should be true if $S^{BH}$ is
considered together with the entropy of a matter outside the
horizon. Black hole must have an intrinsic entropy proportional
to the horizon area.
Otherwise processes like a gravitational collapse
would be at odds with the second law.

Thermodynamics and statistical mechanics of black holes is one of
the most interesting and rapidly developing branches of
black hole physics. In the Einstein theory $S^{BH}$
is a pure geometrical quantity. In real thermodynamical
systems the entropy is the logarithm of the number of microscopic
states corresponding to a given set of macroscopic parameters.
This raises a natural question: Do black holes have microscopic
degrees of freedom whose number is consistent with the
Bekenstein-Hawking entropy?

The main reason why this question is fundamental is because it
goes beyond the black hole physics itself. Its answer  may
give important insight into the as yet mysterious nature of
quantum gravity.

To see this let us start with a simple estimation and consider a
static neutral super-massive black hole with mass $M$ of the order
of $10^9$ solar masses. Such objects are believed to
occur in the centers of certain galaxies. By taking into account
that ${\cal A}=16\pi G^2M^2$  one finds from (\ref{i.01}) that the
entropy of such a black hole is of the order of $10^{95}$.
It is eight orders of magnitude larger than the entropy
of the microwave background
radiation in the visible part of the Universe! What makes
matters even worse is that in the classical theory a black hole
is nothing but an empty space. Thus, an explanation of
the Bekenstein-Hawking entropy is one of those problems which
cannot be solved in classical gravity theory.

Suppose the horizon surface is covered by cells of a Planckian
size $L_{Pl}\sim \sqrt{G}$.  Then, according to (\ref{i.01}), $S^{BH}$
is of the same order as the number of ways to
distribute signs "+" and
"--" over these cells. The appearance of the Planck scale
in this estimate is not an accident. It indicates that a reasonable
resolution of the black hole entropy problem has to be based
on quantum gravity. Moreover, reproduction of $S^{BH}$ by the
methods of statistical mechanics has to be considered as a very
non-trivial test for any candidate theory.

At the present moment the most promising candidate is
believed to be string (D-brane) theory. A successful
statistical-mechanical derivation of $S^{BH}$ for extremal
\cite{StVa:96}-\cite{MaSt:96} and
near-extremal black holes \cite{CaMa:96},\cite{HoSt:96} is
among most important results in this theory during the last decade.
The string computations, however, do not solve the
problem of the Bekenstein-Hawking entropy completely.
They are not universal and, what may be worse,
they are done for models in flat space-times which
are in some sense dual to the string theory on a given black hole
background.
This kind of derivation says nothing about the
real microscopic degrees of freedom responsible for $S^{BH}$ and
where they are located. A review of the string
computations can be found in \cite{Peet:97}--\cite{DMW}.

Another approach to quantum gravity, loop quantum gravity, also
offers an interesting explanation of $S^{BH}$, see \cite{ABCK:97},
\cite{Ashtekar}. Loop quantum gravity is aimed at
a quantum description of the geometry.
The area of a surface in this approach is treated as
an operator. The degeneracy of the eigenvalues of such operators
can be computed.
The suggestion of loop quantum gravity is
that the Bekenstein-Hawking
entropy is related to the degeneracy of eigenvalues of
the area operator which are comparable in magnitude
to the area of the black hole horizon.
However, there is a main oppen issue here
how general relativity, coupled to quantum matter fields,
is recovered from loop quantum
gravity in a suitable low energy limit \cite{Smol:03}.
Till this question is resolved,  it is not clear how to
describe black holes in this approach.

Let us emphasize that the thermodynamics of black holes is determined
only by
the Einstein equations and classical gravitational couplings.
This may indicate
that an understanding of  black hole entropy is possible without
knowing the details of  quantum gravity.
Only a few general features of the fundamental theory may be really
important. If so, the question is: What are those features?

There were two main directions along which this idea
was investigated during the last few years.
One of them was based on the assumption that classical symmetries on a
black hole  background can control the density of states in
quantum gravity and in this way enable one to derive the
entropy of a black hole.

The other direction of research starts from the suggestion that
the origin of the Bekenstein-Hawking entropy is related to the
properties of the physical vacuum in a strong gravitational
field. The Bekenstein-Hawking entropy measures the loss of
information about quantum states hidden inside the horizon.

In this review, we analyze the "pluses" and "minuses"
of the two approaches and show that the two ways of counting
$S^{BH}$ do not necessarily contradict each other.
The review is organized as follows. In section 2 we discuss
two-dimensional conformal theories (CFT) in relation to the problem
of black hole entropy. We start with black holes whose thermodynamical
relations can be interpreted in terms of such CFT's and use these
examples to introduce some properties of the conformal theory.
Special attention is paid to
near-extremal black holes and black holes in anti-de-Sitter (AdS)
gravities. After that we discuss counting
of the Bekenstein-Hawking entropy
by using a near-horizon conformal symmetry.

The relation of  $S^{BH}$ to the entropy of the thermal
atmosphere around a black hole and an entanglement entropy
is discussed in section 3. We argue that in
the most consistent way available at the present moment
this relation can be studied in induced gravity
models. The Einstein gravity in these models is entirely
induced by quantum effects and the underlying theory is free
from the leading ultraviolet divergences.

A possible connection of the two approaches is discussed
in section 4 where we show
how to construct a concrete
representation of the near horizon conformal algebra
in induced gravity. Our conclusions are summarized in the
last section.

One of our purposes is  to present the material in a form
suitable for non-specialists in this field of research, thus
when possible we avoid technical details.
Many interesting topics
related to the black hole entropy problem are not considered here or
discussed briefly.
They can be found in other review works on this subject
(see, for instance,
\cite{BMP}, \cite{Peet:97}--\cite{DMW}, \cite{Jaco:99}--\cite{JP:03}
and further references below).

\section{Black hole entropy and asymptotic symmetries}
\subsection{Black holes which look two-dimensional}
\setcounter{equation}0

Before studying the problem of black hole
entropy, one may ask a simple question:
Are there some familiar physical systems in flat
space-time which are thermodynamically equivalent to a
given black hole?
The equivalence means that the relation between
the mass, temperature and other
parameters of a black hole is the same, after appropriate
identifications, as a relation between the energy, temperature
and other parameters of the corresponding system.
The answer is positive. It turns out, however, that
different black holes
are equivalent to completely different
systems. Moreover, the dimensionalities of the black hole
and the flat space-time do not coincide in general.
Some black holes may have quite complicated thermodynamical
properties\footnote{For example, charged black holes in anti-de Sitter
space-times have a
phase structure similar to that of the van der
Waals-Maxwell liquid-gas systems in a space-time of one-dimension lower
\cite{CEJM}.}, some others are very simple.

Consider a Reissner-Nordstr\"om solution which describes
a charged black hole in Einstein-Maxwell theory
\begin{equation}\label{1.1}
ds^2=-Bdt^2+{dr ^2 \over B}+r^2d\Omega^2~~~.
\end{equation}
Here $d\Omega^2$ is the metric on a unit sphere and
\begin{equation}\label{1.2}
B={1 \over r^2}(r-r_-)(r-r_+)~~~,~~r_{\pm}=m\pm\sqrt{m^2-q^2}~~.
\end{equation}
The parameter $q=Q\sqrt{G}$ is related to the electric charge $Q$ of the
black hole, while $m=MG$, where
$M$ is its mass\footnote{In four dimensions
the Newton constant $G$ (in the system of units we work
in) has the dimensionality $(length)^2$.}.
The radius of the horizon is $r_+$.
The Hawking temperature (\ref{i.01}) of this black hole is
\begin{equation}\label{1.3}
T_H={1 \over 2\pi r_+}\sqrt{m^2-q^2}~~,
\end{equation}
and the Bekenstein-Hawking entropy is
$S^{BH}=\pi r_+^2/G$.

This solution has an interesting property:
the Hawking temperature vanishes in the limit when $m=q$ or
$M=Q M_{Pl}$ where $M_{Pl}=G^{-1/2}$ is the Planck mass.
Such a limiting solution is called an extremal black hole.
Strictly speaking, there are no physical processes
which enable one to turn a charged black hole with $m >q$
to an extremal one\footnote{The reason why
these black holes are different
can be easily seen when going
in (\ref{1.1}) from the Lorentzian to the Euclidean signature.
Then in the $r-t$ plane a non-extremal black hole in a cavity
has the disk topology,
while an extremal black hole looks
like an infinite throat.}.
Macroscopic extremal black holes hardly exist.
These solutions, however, have a theoretical interest
for reasons we discuss later.

We consider now black holes which are "almost extremal"
(or near-extremal) whose mass parameter is
\begin{equation}\label{1.4}
m=q+E~~,~~~E \ll q~~.
\end{equation}
Thermodynamical relations for these objects are very simple.
If we introduce the parameter $\lambda = (2\pi^2 q^3)^{1/2}$ then
\begin{equation}\label{1.5}
T_H\simeq {E^{1/2} \over \lambda} ~~
\end{equation}
and deviations of the mass and the entropy of the black hole
from the extremal values are
\begin{equation}\label{1.6}
E=m-q=\lambda^2 T_H^2~~,~~
S=S^{BH}-{\pi \over G} q^2={2\lambda^2 \over G} T_H~~.
\end{equation}
What can one say about these relations? Consider a gas of some
number of massless
non-interacting scalar fields $\phi_k$ on an interval of  length $b$.
The equations of the fields are
\begin{equation}\label{1.8}
(\partial_t^2-\partial_x^2)\phi_k(t,x)=0~~,~~~
\phi_k(t,0)=\phi_k(t,b)=0~~.
\end{equation}
Suppose that this system is in a
state of thermal equilibrium at some temperature $T$. This
is a one-dimensional analog of an ideal gas of photons in a cavity.
Let us denote the number of field species by $c$.
The free energy of this model is
\begin{equation}\label{1.9}
F(T,L)=cT\sum_n\ln\left(1-e^{-\omega_n/T}\right)~~,
\end{equation}
where the frequencies
of single-particle excitations
$\omega_n=\pi n/b$, $n=1,2,...$, are determined from (\ref{1.8}).
In the thermodynamical limit, $Tb\gg 1$, the series (\ref{1.9})
can be easily calculated
\begin{equation}\label{1.10}
F(T,b)\simeq -{\pi c\over 6}bT^2~~,
\end{equation}
thus, the energy
$E(T,b)$ and the entropy $S(T,b)$ of the system are
\begin{equation}\label{1.7}
E(T,b) \simeq {\pi c\over 6}bT^2~~,~~
S(T,b)\simeq {\pi c\over 3}bT~~.
\end{equation}
For $c=1$ formula for the energy is just an analog of the
Stefan-Boltzmann law. A micro-canonical ensemble is characterized
by the relation
\begin{equation}\label{1.7a}
S=S(E,b)=2\pi\sqrt{{c\over 6}{bE \over \pi}}~~,
\end{equation}
which can be obtained from (\ref{1.7}).
By comparing (\ref{1.7}) with (\ref{1.6}) one can conclude that
thermodynamical properties of a charged black hole near the extremal
limit
are identical to properties of an ideal gas in a flat
two-dimensional space-time.
If we identify in (\ref{1.6}) and
(\ref{1.7}) the temperatures and the entropies, $T_H=T$, $S=S(T,b)$,
then $cb=12\pi q^3L_{Pl}$, where $e$ is the
electric charge of the black hole and $L_{Pl}=\sqrt{G}$ is the
Planck length.

\subsection{Conformal symmetry}

Models (\ref{1.8}) have an important common feature. They possess
a high-level of symmetry which becomes manifest if
the equations
are rewritten in terms of the light-cone coordinates $u=t-x$ and $v=t+x$,
\begin{equation}\label{1.11}
\partial_u\partial_v\phi_k(x)=0~~.
\end{equation}
It is easy to see that equations (\ref{1.11})
are invariant under transformations $u'=f(u)$,
$v'=g(v)$
where $f$ and $g$ are some smooth functions.
These transformations are called conformal
transformations
and the massless 2D quantum field model is an example of
conformal field theory (CFT)\footnote{For a brief
introduction in CFT models see, for example, \cite{Cardy}.}.
In the Euclidean theory,
an analog of these transformations is
$z=f(z')$ and $\bar{z}=\bar{f}(\bar{z}')$ where $z$ and $\bar{z}'$
are coordinates in the complex plane.
Conformal transformations preserve the angle between two vectors
but rescale intervals between neighboring points.

The group of conformal transformations
is an infinite group. To see this it is sufficient to
analyze small transformations of coordinates
$x'^\mu=x^\mu+\delta x^\mu(x)$.
The vector field  $\delta x^\mu$ in the light-cone coordinates
has components
$\delta x^\mu=\zeta^\mu(u)+\bar{\zeta}^\mu(v)$ where
$\zeta^v(u)=\bar{\zeta}^u(v)=0$.
The commutator $[\zeta_1,\zeta_2]$ of two vector fields
\footnote{The commutator, or a Lie bracket, $[\zeta_1,\zeta_2]$ of two
vector fields,
$\zeta_1^\mu$ and $\zeta_2^\mu$, is a vector field with components
$\zeta_3^\mu=\zeta_1^\nu\partial_\nu \zeta_2^\mu-
\zeta_2^\nu\partial_\nu \zeta_1^\mu$.}  is again a vector field,
so one can say that these fields make some algebra with certain
commutation relations. As in the case of the algebra of the rotation
group the algebra of diffeomorphisms can be characterized
by commutation relations in some basis.
Suppose for simplicity that in the model we consider
the fields live on a circle, i.e. instead of the Dirichlet condition
in (\ref{1.8}) we choose a periodic condition
$\phi_k(t,0)=\phi_k(t,b)$. Then one can use Fourier decomposition
for each vector
$$
\zeta^\mu(u)=\sum_n c_n\zeta_n^\mu~~,~~
\bar{\zeta}^\mu(u)=\sum_n d_n\bar{\zeta}_n^\mu~~,
$$
\begin{equation}\label{2.16}
\zeta_n^v(v)={ib \over 2\pi}e^{2\pi i n v /b}~~,~~
\bar{\zeta}_n^u(u)={ib \over 2\pi}e^{2\pi i n u /b}~~,
\end{equation}
where $n$ is an integer and $c_n$, $d_n$ are some constants.
The algebra of these
vector fields has the form
\begin{equation}\label{2.8}
[\zeta_n,\zeta_m]=(n-m)\zeta_{n+m}~~~,~~
[\bar{\zeta}_n,\bar{\zeta}_m]=(n-m)\bar{\zeta}_{n+m}~~,~~
[\zeta_n,\bar{\zeta}_m]=0~~~.
\end{equation}
In fact, one has two commuting sets of generators, each
making an infinite--dimensional algebra
called the Virasoro algebra.

In CFT models the parameter $c$ is called the central
charge. Although
$c$ is not an integer in general, a number of relations,
such, for example, as (\ref{1.7a}) are universal and applicable
for any $c>0$. The central charge is related to an important
property in 2D CFT. The conformal invariance of the classical
equation (\ref{1.11}) is broken in the quantized theory.
This can be seen from the transformation
of the $uu$ or $vv$--components  of the renormalized
stress energy tensor $T_{\mu\nu}=\langle \hat{T}_{\mu\nu}\rangle$
under changes of $u$ and $v$  coordinates. For instance,
under an infinitesimal change
$\delta u=\zeta^u(u)\equiv\varepsilon(u)$, it can be shown that
\begin{equation}\label{2.8a}
\delta T_{uu}(u)=T'_{uu}(u)-T_{uu}(u)= \varepsilon (u) \partial_u T_{uu}(u)+
2\partial_u\varepsilon(u) T_{uu}(u) +{c \over 24\pi}
\partial_u^3\varepsilon(u) +O(\varepsilon^2)
~.
\end{equation}
The term proportional to $\partial_u^3\varepsilon(u)$ is anomalous.
It appears because the renormalization procedure requires
subtracting the divergent part of the stress energy tensor which
is not scale invariant.
This property is  analogous to the chiral anomaly in quantum theory.

Another way to see the conformal anomaly is the following.
In quantum theory the generators of conformal transformations
are some operators acting in the corresponding Fock space \cite{Cardy}.
These operators are expressed in terms of the components
of the stress-energy tensor operator $\hat{T}_{\mu\nu}$.
In this way, in quantum theory each vector $\zeta_n$ ($\bar{\zeta}_m$)
corresponds to some operator $\hat{L}_n$ ($\hat{\bar{L}}_m$)
which form the following algebra:
\begin{equation}\label{2.9}
[\hat{L}_n,\hat{L}_m]=(n-m)\hat{L}_{n+m}+{c \over 12}(n^3-n)\delta_{n+m,0}~~~,
\end{equation}
\begin{equation}\label{2.10}
[\hat{\bar{L}}_n,\hat{\bar{L}}_m]=(n-m)\hat{\bar{L}}_{n+m}+
{c \over 12}(n^3-n)\delta_{n+m,0}~~~,
\end{equation}
$[\hat{L}_n,\hat{\bar{L}}_m]=0$. The brackets $[~~,~~]$ now are
the usual commutators.
Due to the conformal anomaly the quantum algebra
(\ref{2.9}), (\ref{2.10}) differs from the classical one (\ref{2.8})
by the term ${c \over 12}(n^3-n)\delta_{n+m,0}$ which is
called a central extension.

The Hamiltonian operator $\hat{H}$ of the system, which generates
the evolution along the time coordinate $t$ can be expressed
in terms of operators $\hat{L}_0$ and $\hat{\bar{L}}_0$ as
\begin{equation}\label{2.15}
\hat{H}={2\pi \over b}(\hat{L}_0+\hat{\bar{L}}_0)~~~.
\end{equation}
This equation follows from the definition of coordinates
$u$ and $v$, which together with (\ref{2.16}) implies that
$i\partial_t=i\partial_u+i\partial_v={2\pi \over b}
(\zeta_0^\mu+\bar{\zeta}_0^\mu)\partial_\mu$.

In a free quantum field theory the Fock space is constructed
by using creation and annihilation operators.
In the CFT theory there is an alternative way to do this
by using the group algebra  (\ref{2.9}), (\ref{2.10}).
One can do it independently for each copy. Let $|0\rangle$ be the
vacuum vector, such that
\begin{equation}\label{2.17}
\hat{L}_{k}|0\rangle=\hat{\bar{L}}_{k}|0\rangle=0~~~,~~~k=0,1,2,...~~.
\end{equation}
Consider a vector of the Fock space, $|h,\bar{h}\rangle$,
which is  an eigenvector of operators $\hat{L}_0$, $\hat{\bar{L}}_0$
with eigenvalues $h$ and $\bar{h}$, respectively.
This vector will also be an eigenvector of the Hamiltonian
$H$ with energy\footnote{It should be noted that the total energy
of the system is $E+E_0$
where $E_0$ is the energy of vacuum fluctuations. In what follows we assume that
$E$ is large as compared with $E_0$ so $E_0$ can be neglected.}
$E=2\pi(h+\bar{h})/b$. Such a vector
can be constructed by acting on the vacuum with operators
$\hat{L}_{-k}$ and $\hat{\bar{L}}_{-k}$, where $k\geq 1$,
\begin{equation}\label{2.18}
|h,\bar{h}\rangle =
\prod_k (\hat{L}_{-k})^{\alpha_k}\prod_p (\hat{\bar{L}}_{-p})^{\beta_p}|0\rangle~~,
\end{equation}
\begin{equation}\label{2.19}
\sum_k k\alpha_k=h~~,~~~\sum_p p\beta_p=\bar{h}~~.
\end{equation}
The fact that (\ref{2.18}) is an eigenvector of $\hat{L}_0$ and
$\hat{\bar{L}}_0$ can be easily checked by using the Virasoro algebra
(\ref{2.9}), (\ref{2.10}).

As can be seen from (\ref{2.19}) the states
$|h,\bar{h}\rangle$ are degenerate.
Their degeneracy  for large $h,\bar{h}$ can be found exactly.
The degeneracy $D$  corresponding to an eigenvalue $h$ is
\begin{equation}\label{2.20}
\ln D\simeq 2\pi \sqrt{ch \over 6}~~
\end{equation}
(analogously for the degeneracy $\bar{D}$ corresponding
to an eigenvalue $\bar{h}$).
This equation is known as the Cardy formula.
It is applicable to
theories with any central charge $c>0$.

There are different
ways to derive (\ref{2.20}) by using conformal properties.
For our purposes, however, it is more
instructive to see how it follows from results discussed in
section 2.1. Consider
a state of the scalar model with $h=\bar{h}=Eb/4\pi$.
Its degeneracy is related to the entropy $S(E,b)$
of the micro-canonical ensemble with the given energy $E$,
$$
S(E,b)=\ln D+\ln \bar{D}=2\ln{D}~~.
$$
Then the Cardy formula is just the consequence
of the statistical-mechanical relation (\ref{1.7a}).

\subsection{Digression about computations in string theory}

Let us emphasize that in the considered example there is no
apparent relation between a classical black hole and
the quantum model (\ref{1.8}). Suppose, however, that there is an
underlying fundamental theory of quantum gravity able to
provide a statistical explanation of the Bekenstein-Hawking
entropy of a near-extremal black hole. Then the microscopic degrees
of freedom responsible for $S^{BH}$
are to be described by a certain CFT.

String theory provides an explicit example of how this happens
in the case of extremal black holes. These black holes are special
solutions of an effective supergravity theory which is a low-energy
limit of string theory. Typically, solutions in this theory
break the supersymmetry but extremal black holes are invariant
with respect to a part of the supersymmetry transformations.
They are the so-called  BPS solitons and the condition of extremality
$m=q$ is known as a BPS bound. This bound
ensures that the energy of the soliton
receives no corrections in quantum theory.
The Bekenstein-Hawking entropy of an extremal black hole is
\begin{equation}\label{1.12}
S^{BH}={\pi q^2 \over G}= \pi Q^2
\end{equation}
and it does not depend on the gravitational coupling. Therefore, in
gravity theories with different $G$'s
black holes with equal charges $Q$ have equal entropies.

In string theory the gravitational constant $G=g^2l^2$.
It depends
on the string coupling $g$ (the value of the dilaton) and on
the inverse string tension $l$. The parameter $l$ determines
the typical size of a closed string\footnote{In quantum theory
strings with size larger than $l$ give an exponentially small
contribution to the functional integral.}.  Because $S^{BH}$
in (\ref{1.12})
does not depend on $G$ one can vary the string coupling $g$
without changing the entropy of an extremal black hole.
Note, however, that the size of the black hole associated
with the horizon radius $r_+$ depends on the gravitational
constant, $r_+=MG=Q\sqrt{G}=Qgl$.

One can consider two limits. In the limit of weak coupling $g$,
the horizon radius can be much smaller than the string size,
$r_+\ll l$. In this limit, instead of a black hole
one has a dual object,
a point particle in a flat space.
A black hole is formed in the limit of strong coupling
when $r_+\gg l$. However, when one increases the coupling
and goes from the weak limit to the strong one $S^{BH}$
does not change. This means, that instead of doing calculations
of the entropy on the black hole background one can consider a dual
theory
in a flat space-time, which is much easier.
It turns out that the dual theory is a CFT similar to what we described
above and counting its states gives the correct value of $S^{BH}$
\cite{StVa:96}.

No doubt, this result is  important
but it is not quite satisfactory. Since computations are done
in a  dual theory the physical nature and the location
of the black hole degrees of freedom remain unknown.
It is also not clear whether it is possible to extend this analysis
to be applicable to any black hole.

\subsection{Anti-de Sitter black holes}

Another interesting example where thermodynamical relations of
black holes are equivalent to relations emerging in CFT models
appears in the three dimensional (3D) gravity theory with a negative
cosmological
constant $\Lambda=-l^{-2}$. The theory is described by the action
\begin{equation}\label{2.0}
I={1 \over 16\pi G_3}\int\sqrt{-g}d^3x\left(R+{2 \over l^2}\right)~~,
\end{equation}
where $R$ is the scalar curvature and $G_3$
is the 3D gravitational coupling (note that
$G_3$ has the dimension of length).
The important feature of gravity in three dimensions
is that if $\Lambda=0$ the space-time geometry
is always locally flat. A point mass in such theory does not have
a gravitational potential but changes global properties of the
geometry around itself.
Another feature of 3D gravity is the absence of gravitons.

One of the solutions of  3D gravity
with nonzero cosmological constant is anti-de Sitter $AdS_3$
space\footnote{For this reason such theories are also called
AdS gravities.}
\begin{equation}\label{2.1}
ds^2=-\left(1+{r^2 \over l^2}\right)dt^2+
\left(1+{r^2 \over l^2}\right)^{-1}dr^2+r^2d\varphi^2~~,
\end{equation}
where $0\leq \varphi \leq 2\pi$. This space
has a constant negative
curvature and can be defined as the surface $x^2+y^2-z^2-w^2=-l^2$
in a flat 4D space with metric $ds^2=dx^2+dy^2-dz^2-dw^2$.
It is denoted as $AdS_3$.

There are also black hole solutions in this theory
discovered in \cite{BTZ} and called BTZ black holes after
the authors. The metric of a BTZ black hole is simple
\begin{equation}\label{2.2}
ds^2=-{r^2-r_+^2 \over l^2}dt^2+
\left({r^2-r_+^2 \over l^2}\right)^{-1}dr^2+r^2d\varphi^2~~.
\end{equation}
The horizon is located at  $r=r_+$ and is a circle.
The area of the horizon is the length $2\pi r_+$ of this circle.
Space
(\ref{2.2}) has locally the same geometry as (\ref{2.1})
but differs from it by global properties.
We denote the black hole space-time by ${\cal M}_3$
to distinguish it from $AdS_3$.

The mass $M$ of the BTZ black hole is defined as $M=r_+^2/(8l^2G_3)$.
Thus, the relation between the Bekenstein-Hawking entropy
and the mass is
\begin{equation}\label{2.3}
S^{BH}={2\pi r_+ \over 4G_3}=2\pi\sqrt{{l^2 M\over 2G_3}}~~.
\end{equation}
This formula has the same form as Eq. (\ref{1.7a}) for the
entropy of a CFT. To make the correspondence more precise
note that for a black hole in AdS gravity the curvature radius $l$
plays the role of a "size" of the black hole\footnote{To explain this
analogy we take a different example and
consider a Schwarzschild black hole in the Einstein theory.
It is known that this black hole is thermodynamically unstable.
There are two ways to solve this problem: to place the black
hole in a spherical cavity of a certain radius \cite{York:86}
or to introduce a negative cosmological constant.}.
Thus, $l$ in (\ref{2.3})
is analogous to parameter $b$ in (\ref{1.7a}).
If we identify the entropy and the mass of the black hole
with the entropy and the energy of the CFT then
the central charge $c$ of this theory has to be proportional
to $l/G_3$. In fact, this is a very good guess:  the correct
value of the central charge is
\begin{equation}\label{2.4}
c={3l \over 2G_3}~~.
\end{equation}
A remarkable property of  BTZ black holes is
that the corresponding conformal
group is realized not in a dual theory, as in the case of
extremal black holes, but as group of the asymptotic transformations
of the physical space-time background.

As one case see, at large $r$ the BTZ metric (\ref{2.2}) behaves
like the
$AdS_3$ metric (\ref{2.1}).
One says that BTZ geometry is asymptotically $AdS$.
There is a group of coordinate transformations
$\delta x^\mu=\zeta^\mu(x)$ which preserves
this asymptotic structure.
At large $r$ the diffeomorphism
vector fields are \cite{Stro:97}:
\begin{equation}\label{2.5}
\zeta^t=l(T^+ +T^-)+{l^3 \over 2r^2}(\partial_+^2T^+
+\partial_-^2T^-)+O(r^{-4})~~,
\end{equation}
\begin{equation}\label{2.6}
\zeta^\varphi=(T^+ -T^-)-{l^2 \over 2r^2}(\partial_+^2T^-
-\partial_-^2T^-)+O(r^{-4})~~,
\end{equation}
\begin{equation}\label{2.7}
\zeta^r=-\frac 12 r(\partial_+T^+ +\partial_+T^+)+O(r^{-1})~~~,
\end{equation}
where $\partial_{\pm}=l\partial_t \pm \partial_\varphi$,
$T^{+}$ is a function of single variable
$t/l+\varphi$ while $T^{-}$ is a function of $t/l-\varphi$.

It is not difficult to check that the
commutator
of vector fields $\zeta_1$ and $\zeta_2$ which have
the asymptotic behavior
(\ref{2.5})--(\ref{2.7}) with functions $T_1^{\pm}$ and $T_2^{\pm}$,
respectively,
is a vector field $\zeta_3$ which has the same asymptotic
behavior with
$T_3^{\pm}=T_1^{\pm}\partial_{\pm}T_2^{\pm}-
T_2^{\pm}\partial_{\pm}T_1^{\pm}$. Thus, one can say that
generators of these diffeomorphisms form a closed algebra.
There is a natural choice of basis in this algebra,
$\zeta_n$, $\bar{\zeta}_n$, which is singled out by
the the following restrictions on the corresponding functions:
\begin{equation}\label{2.13}
T_n^{+}=\frac i2 e^{i n(t/l+\varphi)}~,~~ T_n^{-}=0~,~~
\bar{T}_n^{+}=0~,~~
\bar{T}_n^{-}=\frac i2 e^{i n(t/l-\varphi)}~~,
\end{equation}
where $n$ is an integer.
The algebra of these
vector fields is given by relations (\ref{2.8}) and
therefore the group of asymptotic transformations
(\ref{2.5})--(\ref{2.7}) is the conformal group discussed
in section 2.2.

We have pointed out that the representation of the conformal algebra
in quantum theory acquires a central extension due to
the conformal anomaly, see (\ref{2.9}), (\ref{2.10}).
It is an interesting fact that the Virasoro algebras with central
extension also appear in classical theory, as
happens for example, in Liouville theory \cite{HJ}.
In a classical theory the symmetries can be realized in a phase
space.
If there is a symmetry group its algebra is represented by
relations
where the Poisson bracket plays the role of the commutator.
In the classical gravity (\ref{2.0})
the generators of diffeomorphisms $\delta x^\mu=\zeta^\mu(x)$
have the following structure
\begin{equation}\label{2.11}
H[\zeta]=\int_{\Sigma_t}\zeta^\mu \phi_\mu d\Sigma+J[\zeta]~~.
\end{equation}
The integral goes over a space-like hyper-surface $\Sigma_t$
of constant time $t$. Quantities
$\phi_\mu$ and $J[\zeta]$ depend on canonical coordinates, $g_{ij}$, and
momenta, $\pi_{ij}$, which are defined from
the Lagrangian by standard methods.

To avoid the divergences in the theory at large radii $r$
one has to restrict the integration over
$\Sigma_t$ by some large upper bound $r\leq r_0$
($r_0$ can be taken to infinity in the last stage of the computation).
The last term
$J[\zeta]$ in (\ref{2.11}) is a surface term defined at $r_0$.
Quantities $J[\zeta]$ are introduced to ensure a canonical form for
variations of $H[\zeta]$,
\begin{equation}\label{2.12}
\delta H[\zeta]=\int_{\Sigma_t}(A^{ij}\delta g_{ij}+B^{ij}\delta
\pi_{ij})~~~.
\end{equation}
The form of $J[\zeta]$ in general depends on the boundary
conditions at $r_0$.

The equations $\phi_\mu=0$ are constraints analogous to the Gauss law
$\nabla E-\rho=0$ in electrodynamics. Thus, when
the equations of motion are satisfied,
$H[\zeta]$ reduce to pure surface terms.
For this reason, in particular, the energy of the system
which is associated with the generator of translations
$i\partial_t=i\zeta^\mu_{(t)}\partial_\mu$
along the
Killing time is non-trivial because of the presence of the surface
term $J$.
The on-shell value of $H[i\zeta_{(t)}]$ is defined
so that to coincide with mass $M$ of the BTZ black hole, Eq. (\ref{2.2}).

One can define generators $L_n=H[\zeta_n]$,
$\bar{L}_n=H[\bar{\zeta}_n]$
corresponding
to the particular set of diffeomorphism vectors having
the asymptotic form (\ref{2.5})--(\ref{2.7}).
Their canonical
commutation relations were investigated by Brown and Henneaux
\cite{BrHe} who found that
$L_n$, $\bar{L}_n$ form a Virasoro algebra
isomorphic to (\ref{2.9}), (\ref{2.10}), where the constant
$c$ is the central extension given by (\ref{2.4}).

\bigskip

As follows from (\ref{2.5})--(\ref{2.13}),
the generator of time translations is represented as
$i\partial_t=i(\partial_+ +\partial_-)/2l=l^{-1}
(\zeta_0^\mu+\bar{\zeta}_0^\mu)\partial_\mu$.
Therefore, one has the following relation between the
energy and the Virasoro generators
\begin{equation}\label{2.14}
H[i\zeta_{(t)}]=\frac 1l(H[\zeta_0]+H[\bar{\zeta}_0])=
\frac 1l(L_0+\bar{L}_0)~~.
\end{equation}
This equation is analogous to relation (\ref{2.15})
discussed in section 2.1. Now, however, (\ref{2.14}) is
a classical quantity defined on a phase space.
Suppose that modulo the
equations of motion $L_0=h$, $\bar{L}_0=\bar{h}$.
If the energy of the system coincides with the mass of the black hole,
then, by the symmetry, $h=\bar{h}=Ml/2$.

Do these observations say something about the entropy of the BTZ
black hole? It is a well-known fact that there is a correspondence
between the Poisson brackets in  classical
mechanics and commutators of operators in quantum theory.
By taking this into account one can make the following suggestions:

\bigskip

i) There is a quantum gravity theory on $AdS$ such that  physical
states of this theory yield a representation of the
Virasoro algebra related to asymptotic symmetries;
classical generators $L_n$, $\bar{L}_n$
correspond to operators $\hat{L}_n$, $\hat{\bar{L}}_n$
in the quantum gravity.

ii) The central charge $c$ of the Virasoro algebra in quantum gravity
coincides with the central charge (\ref{2.4}) of the classical theory.

iii) A quantum state for which the
operators $\hat{L}_0$, $\hat{\bar{L}}_0$ have eigenvalues
$h=\bar{h}$ corresponds to a static BTZ black hole of mass $M=2h/l$.

\bigskip

As follows from (iii) the mass of a black hole cannot be arbitrary
but takes some discrete values
which can be derived by using the
commutation relations of the Virasoro algebra.
The spacing between two levels is determined by the inverse radius
$l^{-1}$ of $AdS_3$. If $M$ is comparable to $l$ the black hole is
essentially a quantum object.
Note that the semi-classical regime of quantum gravity theory also
requires that ${\cal R}\gg L_{Pl}$, where ${\cal R}$ is
characteristic radius of the space-time curvature,
$L_{Pl}=G_3$ is the Planck length in three dimensions.
The geometry of the BTZ black hole has two
such radii, $l$ and $r_+=l\sqrt{8G_3 M}$.
Thus, the semi-classical
limit requires that $l\gg G_3$ and $M \gg G_3/l^2$. The  first condition
imposes a restriction on the central charge
(\ref{2.4}), $c\gg 1$. The second condition holds if
the black hole mass is larger than the Planck mass, $M\gg G_3^{-1}$.
Both conditions then imply that $M\gg l^{-1}$. This means that
the spectrum of a semi-classical black hole can be considered as continuous.

The classical black hole is a highly degenerate object.
The degeneracy $D$, $\bar{D}$ of operators
$\hat{L}_n$, $\hat{\bar{L}}_n$ ,
can be found by using the Cardy formula  (\ref{2.20}). The total
degeneracy is
\begin{equation}\label{2.21}
\ln D+\ln\bar{D}=2\ln D\simeq 4\pi \sqrt{ch \over 6}=
2\pi\sqrt{{l^2 M\over 2G_3}}
~~,
\end{equation}
which is exactly the Bekenstein-Hawking entropy (\ref{2.3})
of the BTZ black hole with mass $M$. The above derivation
of (\ref{2.21}) was first given by Strominger \cite{Stro:97}.
The result can be generalized to the case of a rotating black hole
whose state has additional number, a spin.
One can also investigate along these line black holes
in a 2D $AdS$ gravity \cite{CaMi:a}, \cite{CaMi:b}.

There are a number of technical questions in these derivations
which can be addressed \cite{Carl:98} but can hardly be resolved
without more detailed information about quantum gravity on $AdS_3$.
These questions are related to assumptions (ii) and (iii)
which may not hold
because quantum effects change classical quantities and these changes
are not always small.
However, if the assumptions
(i)--(iii) are adopted one gets a definite answer to
the question formulated in the title
of this review. Note that
those few properties of the quantum gravity theory
relevant for the entropy counting are determined only by
the low-energy constants, $l$ and $G_3$. These constants
define the energy spectrum completely, all one needs to know!
Therefore, the BTZ black hole certainly
sets an example where one can understand black hole entropy
without knowing much about quantum gravity.

Let us emphasize that the above discussion concerns
black holes in $AdS_3$ gravity without matter.
Introduction of matter fields makes
such a derivation of the entropy
impossible in general. Discussion of 3D
black holes with matter fields and further references
can be found in \cite{Park:04a}.

\subsection{AdS/CFT correspondence}

Relation (\ref{2.3}) can be used to find the Hawking temperature of the
BTZ black hole $T_H=(dS^{BH}/dM)^{-1}=(2G_3 M/\pi^2 l^2)^{1/2}$.
If the black hole is considered as a canonical ensemble
one can introduce its free energy, $F_{bh}(T,b)$, via
the standard thermodynamical relation
\begin{equation}\label{2.22}
F_{bh}(T,b)=M-TS^{BH}=-M=-{\pi c\over 6} bT^2~~~,~~~b=2\pi l~~,~~T=T_H
~~,
\end{equation}
where $c$ is given by (\ref{2.4}). It is instructive to compare
this result with the free energy (\ref{1.10}) of the model
discussed in section 2.1 and see that $F_{bh}(T,b)$
is
equivalent to the free energy of  $c$ quantum fields living
on a circle of the length $b=2\pi l$ so that one can
write\footnote{Let us recall that (\ref{1.10}) is applicable in the
thermodynamical limit $TL\gg 1$ which requires
that the black hole is classical, $M\gg G_3^{-1}$.}
\begin{equation}\label{2.23}
F_{bh}(T,b)=F_{CFT}(T,b)
~~.
\end{equation}
This result could be expected from the previous
discussion.
Equation (\ref{2.23}) relates classical and quantum quantities.
The conformal theory lives on a flat space-time
$\tilde{{\cal M}}_2$
which is one dimension lower than ${\cal M}_3$.
The metric of $\tilde{{\cal M}}_2$ is $d\tilde{l}^2=-dt^2+l^2d\varphi^2$.
On the other hand, the metrics of
constant-radius hypersurfaces of ${\cal M}_3$
at large $r$  have the form $dl^2\sim (r/l)^2 d\tilde{l}^2$.
Thus, up to a scale,
factor $\tilde{{\cal M}}_2$ has the same geometry
as asymptotically distant sections $r=const$ of ${\cal M}_3$.
In this sense $\tilde{{\cal M}}_2$ can be called
the asymptotic infinity of  ${\cal M}_3$ or an asymptotic
boundary\footnote{There is a conformal transformation
of the $AdS_3$ and BTZ metrics which maps these spaces
to spatially compact space-times such that in the transformed
metrics the surface $r=\infty$ is located at a finite distance
and defines a boundary, see  \cite{Witt} for the details.}.

It can be shown \cite{BTZ} that $F_{bh}(T,b)$
in (\ref{2.23})
can be obtained
from the classical gravitational action (\ref{2.0})
on the black hole background ${\cal M}_3$.
If one had a quantum gravity on $AdS_3$ the semi-classical
limit of this theory in the black hole sector would be
given by $F_{bh}(T,b)$.
Therefore,
a semi-classical limit of quantum gravity theory on $AdS_3$
is  determined by a conformal
field theory defined at the asymptotic infinity of the bulk
space-time. This property is known as the $AdS/CFT$ correspondence.

There are arguments \cite{Mald},\cite{Witt},\cite{GKP} based on
string theory that the $AdS/CFT$  correspondence also holds for
higher-dimensional $AdS$ gravities.
For a $D$--dimensional
$AdS$ background ${\cal M}_D$ the asymptotic boundary
is a $D-1$ dimensional space-time $\tilde{{\cal M}}_{D-1}$.
The boundary theory living on $\tilde{{\cal M}}_{D-1}$ is
a quantum conformal theory $CFT_{D-1}$.
It should be emphasized that if $D>3$
the properties of the boundary theory cannot be inferred
from the asymptotic symmetries of the background space-time.
The asymptotic symmetry in this case is just the
anti-de Sitter group which is finite-dimensional and does not admit
non-trivial central extensions  in general \cite{BrHe}.
To get the energy spectrum of the CFT more data about
the gauge group of the theory,
its coupling constants and others are required.
String theory provides an example how these data
can be related to the properties of the fundamental gravity
theory\footnote{The AdS/CFT correspondence in string theory
is formulated as follows \cite{Mald}: type IIB string theory on $AdS_5\times S_5$
is dual to ${\cal N}=4$, $D=3+1$ super-Yang-Mills theory with $SU(N)$ group.
Coupling constant, $g_{YM}$, in this theory is related to string coupling constant
$g_{st}$ ($g^2_{YM}\sim g_{st}$), $N$ equals to five-form flux on $S^5$ ($N$
is supposed to be large). Radius of curvature of the background  is
proportional to $(g^2_{YM}N)^{1/4}$.}.

What is important, however, is that in these examples
the characteristics of the boundary
$CFT$ are expressed in terms
of the low-energy parameters. For example, for five-dimensional AdS gravity
the effective number of degrees of freedom of the
corresponding CFT (an analog of the central charge
(\ref{2.4})) is
proportional to $l^3/G_5$ where $l$ is the $AdS$ radius and $G_5$ is the
Newton constant.
In this regard,
the higher-dimensional case
is similar to the BTZ black hole.
It supports the idea that
by using the low-energy parameters the entropy
of higher-dimensional $AdS$ black holes can be reproduced
by the methods of statistical-mechanics without
knowing the details of quantum gravity theory.

A final remark is in order.
If the $AdS/CFT$ correspondence holds, the
information about bulk degrees of freedom in $AdS$
is encoded into a dual boundary theory.
This is an example of how a "holographic principle"
first formulated by 't Hooft \cite{Hooft:93} (see also \cite{Suss:94}
and the review \cite{Bousso}) is realized.
This property does not explain what are the bulk degrees of freedom
and where are they located but it may help to resolve
other problems. For instance,
since the boundary theory is unitary so should be the
process of black hole evaporation.

\subsection{Near-horizon conformal symmetry}

The arguments based on the $AdS/CFT$ correspondence are
not universal because they are restricted
to gravity theories with a negative cosmological
constant. They are not applicable to the most interesting case
of asymptotically flat black hole space-times.

It is easy to understand where the difficulty comes from.
The problem of the Bekenstein-Hawking entropy $S^{BH}$
is related to the physics
near the black hole horizon. The value of the entropy, the
temperature of the Hawking radiation and properties
of the spectrum of the radiation (the so called gray-body factors)
are determined by the space-time geometry near the horizon.
These facts strongly suggest that a universal
approach to $S^{BH}$ should be related to the near
horizon region\footnote{This may not be necessarily true because
the black hole entropy is a global quantity \cite{HaHu:98}.}
rather than to spatial infinity.

It is natural to ask whether one can derive the Bekenstein-Hawking entropy
by applying the so far successful arguments based on a symmetry group
to the region
near the horizon. The first attempts in this direction were
made by Carlip \cite{Carl:99a} and Solodukhin
\cite{Solo:99} and then continued in large number of
publications by other authors \cite{DGW}--\cite{KKP:04}.
We will not attempt to
describe these works here in full detail.
This would require us to go into many technical questions
which are not completely
resolved\footnote{The latest account of these results
and references can be found in \cite{KKP:04}.}.
Also there is no unique point of view as to how this approach
should be realized.
We focus on
some general features related to the formulation of
this problem.

In the region near the horizon the black hole metric takes a
simple form
\begin{equation}\label{2.24}
ds^2=-\kappa^2 \rho^2 dt^2+d\rho^2+d\sigma^2
~~.
\end{equation}
The horizon is located at $\rho=0$.
The coordinate $\rho$ is the proper distance from a point
to the horizon and
$d\sigma^2$ is the metric on the horizon
surface. Asymptotically (\ref{2.24}) is valid for non-extremal
black holes which have a non-vanishing surface
gravity constant $\kappa$ (and, hence, a non-zero Hawking temperature
$T_H$, see (\ref{i.01})). Formula (\ref{2.24}) is called the Rindler
approximation. If $d\sigma^2=dx^2+dy^2$ is a flat metric,
(\ref{2.24}) is the metric in Minkowski space written in
Rindler coordinates.
An observer moving along the trajectory $\rho=const$
has acceleration $1/\rho$.

By using the BTZ black hole as an example one has to look for a
relevant group of coordinate transformations which preserves
this form of the metric and is isomorphic to the conformal group.
This can be done in many ways, but a universal approach
should be applicable to black holes in different gravity
theories. In particular, it must work in two-dimensional gravities
where the black hole horizon is a point\footnote{More precisely,
the cross-section of the black hole horizon and a constant time
hyper-surface in two-dimensional black holes is a point.}
and $d\sigma^2=0$ in (\ref{2.24}). Thus, it is natural
to identify the conformal group with coordinate transformations
in the $t-\rho$
plane,  as was first proposed in \cite{Solo:99}.
In arbitrary dimensions this is a
two-dimensional plane $\cal G$ orthogonal to the horizon surface.
Let us denote its metric as $d\gamma^2$.
In the light-cone coordinates
\begin{equation}\label{2.25}
d\gamma^2=-\kappa^2 \rho^2 dt^2+d\rho^2=-\kappa^2 \rho^2 dudv
~~,
\end{equation}
\begin{equation}\label{2.26}
u=t-x~~,~~v=t+x~~,~~x={1 \over \kappa}\ln \rho~~.
\end{equation}
The coordinate transformation which lead to the conformal group
are those discussed in section 2.2, i.e. $u'=f(u)$, $v'=g(v)$.
Suppose this choice of transformations is correct.
Can it be used to reproduce the Bekenstein-Hawking entropy?
To answer this question note that
there are several key distinctions
between the near-horizon approach and the approach used in
the case of the BTZ black hole.

i) The thermodynamical relations for a black hole
in the near-horizon region do not look like relations of a 2D CFT.
An observer at rest with respect to the black hole horizon
measures a temperature of the Hawking radiation $T$ which
differs from the Hawking temperature $T_H$ by a blue-shift factor,
$T=T_H/\sqrt{B}$ where $B$ is related to the time-component
of the metric
(it is the modulus of norm of the Killing vector $\partial_t$).
Near the horizon $B\simeq\kappa^2 \rho^2$. Thus, according
to (\ref{i.01}) the local temperature is
$T\simeq 1/(2\pi \rho)$.  It is determined only
by the acceleration of the observer and does not depend on
black hole parameters. According to York \cite{York:86},
if the black hole is placed in a cavity
its temperature is defined as a local-temperature $T$ on
the boundary of the cavity .
The black hole is characterized by an energy
$E$ which should be consistent with the first law of thermodynamics.
For instance, if the radius of the cavity is fixed, $dE=TdS^{BH}$.
However, when the boundary of the cavity is moved close to
the horizon, $T$ becomes a free parameter which means that
$E=TS^{BH}$ up to an additive constant.
Therefore, $S^{BH}\sim E$ and this relation
differs from (\ref{1.7a})\footnote{It is interesting to note that
such a relation between the energy and the entropy is typical
for string theory
where the degeneracy of a level with
the energy $E$ is proportional to $e^{E}$ \cite{Horo:97}.}.

ii) Approximation (\ref{2.25}) leaves
only two parameters:
the surface area of the horizon ${\cal A}$ and the Planck length
$L_{Pl}$ (defined by the Newton coupling constant in the given
theory). For a BTZ black hole there is an extra parameter,
the $AdS$ radius $l$, which determines
the spacing between the energy levels.

iii) The boundary conformal theory in the BTZ case is given on a circle
of length $2\pi l$. Contrary to this in the near-horizon
approach the light-cone coordinates are not compact. Therefore,
to have a discrete basis of generators of the conformal algebra
$L_n$, $\bar{L}_n$ one has to impose some boundary conditions
on diffeomorphisms in the $t-\rho$ plane and introduce an extra
parameter $b$, the size of the space where
the CFT theory is defined. This scale should determine the
spacing in the energy spectrum.

iv) Suppose that in the gravity theory
the algebra of generators $H[\zeta_n]$ of the diffeomorphisms
in the $t-\rho$ plane is a Virasoro algebra (\ref{2.9})
with a central charge  $c$.
What is the value of  $c$ in this theory?
The black hole is identified with a certain
quantum state with the energy $E$.
To relate the Bekenstein-Hawking
entropy to the degeneracy of the energy level $E$ one has to use
(\ref{1.7a}) and condition that $E \sim S^{BH}$. This
requires that $c\sim S^{BH}$.
The precise value of $c$ is fixed when $b$ is fixed.

\bigskip

The derivation of the Bekenstein-Hawking entropy
along these lines was
given in \cite{Carl:99a},\cite{Solo:99} and in
subsequent publications. It should be noted that these derivations
used a single copy of the Virasoro algebra and the conformal
transformations were not necessarily
related to transformations of coordinates $u$ and $v$. However,
all these works despite technical differences
had the basic features described above.

An attractive feature of the near-horizon approach is
its universality and a certain hope
to explain the black hole entropy without relying on the details
of quantum gravity theory. But is this hope justified?

One of the problems is that the central charge $c$
in the boundary CFT is proportional to the area of the black hole horizon.
This  means that $c$ depends on the background, a property which does not
look natural.
Let us recall that
the central charge in
$AdS$ gravities is a combination of the fundamental
constants, see (\ref{2.4}). For this reason, the $AdS/CFT$ correspondence
enables one to consider different backgrounds
(for example, a black hole and a pure anti-de Sitter space)
as different quantum states of the same boundary CFT.
As a result, black hole evaporation is equivalent
to a time evolution in some CFT. There should be
no loss of information in this process.
In contrast to this in the near-horizon approach black holes
with different masses correspond to states in {\it different} CFT's.
The evaporation of a black hole is an evolution in a
space of  theories and it is not restricted by
requirements of unitarity.

The other problem is the choice of the boundary conditions
at the horizon and fixing the central charge.
It is clear that the Rindler approximation is not enough
for this purpose. Perhaps, other characteristics of the
gravitational field in the vicinity of the horizon may help
to define the CFT completely. Some work in this direction can be found
in
\cite{Solo:03}. On the other hand,
going beyond the Rindler approximation
certainly puts at risk the universality of the method.

The problem may be even more serious: to fix the
boundary CFT one needs to know those details of the quantum
gravity theory which are not available at low energies.
The approach based on the near-horizon symmetry gives at best
a statistical
{\it representation} of $S^{BH}$. It implies the existence of the
corresponding micro-states but does not prove it.
Note that the $AdS/CFT$ correspondence is supported by
computations in string theory  \cite{Witt}--\cite{GKP}
while approach \cite{Carl:99a},\cite{Solo:99} does not have such support
so far.

Are there any examples of a microscopic realization of the near-horizon
symmetry and what can one  learn from them?
This will be discussed in the second part of the paper.

\section{Black hole entropy as a property of the physical vacuum}
\setcounter{equation}0
\subsection{Thermal atmosphere and entanglement entropy}

We now turn to another approach where the origin of the black hole
entropy is related to the properties of the physical vacuum
in strong gravitational
fields. There are always zero-point fluctuations
of physical fields in a vacuum state. An observer, who is at rest
with respect to the horizon sees these vacuum excitations
as a thermal atmosphere around a black hole
\cite{ZuTh:85}--\cite{MuIs:98}. The first attempts to relate
the Bekenstein-Hawking entropy to the thermal atmosphere
were made by Thorne and Zurek \cite{ZuTh:85}  and by 't Hooft
\cite{Hooft:85}.

Let us calculate, as an example,
the entropy $S$ of a quantum scalar field around a static
black hole. First note that near the horizon the local temperature
is $T=1/(2\pi \rho)$ and it grows indefinitely when the horizon is
approached ($\rho$ goes to zero).
Thus, one can use the high-temperature
asymptotic form of the free energy in a gravitational field.
This asymptotic form is well known.
In four-dimensional
static space-time the free-energy
is\footnote{Finite-temperature quantum field theory
in gravitational backgrounds including the case of black hole
backgrounds is discussed in
\cite{FrFu:98}--\cite{Furs:01b}.}
\begin{equation}\label{3.1}
F(\beta)\simeq -{\pi^2 \over 90}\int \sqrt{-g} T^4 d^3x~~.
\end{equation}
Here $g$ is the determinant of the metric, $T=\beta^{-1}/\sqrt{|g_{00}|}$
is the local temperature, $g_{00}$ is the time-component of the metric.
In asymptotically flat space-times, like a Schwarzschild black hole,
$\beta^{-1}$ is the temperature measured by an observer at infinity.
For our purposes evaluation of (\ref{3.1})
can be done by using the Rindler approximation (\ref{2.24}). By taking into
account that
$g=\kappa\rho$, $d^3x=d\rho d^2\sigma$ one can see that the integral in
(\ref{3.1}) diverges. Let us
introduce a cutoff at some small distance
$\epsilon$ near the horizon. The leading contribution to
entropy can be found by using the standard statistical-mechanical
definition
\begin{equation}\label{3.2}
S=\beta^2 {\partial F(\beta) \over \partial \beta}
\simeq {1 \over 360 \pi \epsilon^2}{\cal A}~~.
\end{equation}
The quantum field is supposed to be in  thermal
equilibrium with the black hole. This is possible when the temperature
coincides with the temperature of the Hawking radiation.
Thus, when the derivative is taken one puts $\beta=\kappa/(2\pi)$
and gets the right-hand side of (\ref{3.2}).
The quantity ${\cal A}$ is
the integral $\int d^2\sigma$ which is the surface area of the horizon.

It is natural to assume \cite{Hooft:85} that
the cutoff parameter is comparable to the
Planck length, $\epsilon\simeq \sqrt{G}$.
Then $S$ in (\ref{3.2}) has the same
order of magnitude as the Bekenstein-Hawking entropy $S^{BH}$
of a black hole.

One may wonder how the entropy can  be
related to properties of the vacuum.
The explanation is that static observers
near a black hole horizon perceive the vacuum as a mixed state.
This happens
because they cannot do measurements inside
the horizon.  There is a non-trivial density matrix $\hat{\rho}$ which
appears because in a local
quantum field theory "observable" and "non-observable" vacuum fluctuations
are correlated or entangled at the horizon.
There is an information loss which can be
quantified by some {\it entanglement}
entropy $S_{ent}=-\mbox{Tr}\hat{\rho}\ln \hat{\rho}$.
A remarkable property of black holes is that the entanglement
entropy coincides with the entropy of the thermal
atmosphere\footnote{The fact that  $S_{ent}$ is proportional
to the horizon area and can be related to the Bekenstein-Hawking
entropy was first pointed out in \cite{BKLS:86}--\cite{FrNo:93}.
This entropy was then studied in
\cite{LaWi:95}--\cite{Muko:98}.}
because
$\hat{\rho}$ is a thermal density matrix
\cite{Israel:76}--\cite{FR}.

Can $S$ (or $S_{ent}$) be the source of the Bekenstein-Hawking entropy?
To answer this question one has to resolve the following problems:

i) $S$ depends on the cutoff $\epsilon$. Therefore,
there must be some natural explanation why the cutoff is adjusted
so that $S=S^{BH}$.

ii) In the general case $S$ receives contributions from all
fields present in the Nature. It depends on the total
number of fields and their spins. However, $S^{BH}$ does not have
such dependence.

Before we consider these problems one more
property of the thermal entropy has to be discussed.
Introduction of the cutoff $\epsilon$ means that a quantum field
cannot propagate on the entire black hole background. It cannot leak
inside the horizon because of some artificial ("brick wall")
boundary conditions.
It should be emphasized that the
horizon is not a boundary and there can be no conditions
in this region but regularity.

There are other regularizations of the integral (\ref{3.1})
consistent with this property. For instance,
one way to get rid of the divergences would be to use dimensional
regularization. In a $D$ dimensional space the integral
(\ref{3.1}) depends on $T^D$ and converges if $D$ is extrapolated
to the region $D<2$.

One can also use, as was suggested in \cite{DLM:95},
a Pauli-Villars (PV) regularization.
In
this method for each physical field, one introduces   $5$
additional
auxiliary fields: $2$ fields with
masses $M_k$ which have the same statistics as the original field and  $3$
fields with masses $M_r'$ which have the wrong statistics.
The masses can be chosen as follows
$M_{1,2}=\sqrt{3\mu^2+m^2}$, $M'_{1,2}=\sqrt{\mu^2+m^2}$,
$M'_3=\sqrt{4\mu^2+m^2}$ where
$\mu^2$ plays the role of a regularization parameter.
The leading part of the entropy of each PV field
is given by formula (\ref{3.2}) with the only difference that
fields with the wrong statistics give negative contributions to
the total entropy. For this reason the leading divergence is canceled.
To find
the entropy one has to use next-to-leading terms in the
high-temperature asymptotic expressions
(see \cite{FrFu:98} for the details).
The final result in the limit of large $\mu$ is
\begin{equation}\label{3.3}
S=S(\mu) \simeq
{\lambda \over 48\pi} \mu^2 {\cal A} ~~~,
\end{equation}
where $\lambda=\ln{729 \over 256}$. The divergence in $S$ in the PV
method appears in the limit of infinitely heavy PV fields.

Both dimensional and PV regularizations are used in quantum field
theory to regularize ultraviolet divergences in Feynman
diagrams. The fact
they can be used for the entropy indicates that the divergences
near the horizon may be related to the ultraviolet divergences.
This is in fact true and, as was first suggested by Susskind and Uglum
\cite{SuUg:94} and Callan and Wilczek \cite{CaWi:94},
these divergences can be removed
by the standard renormalization of the Newton constant.

\subsection{Entanglement entropy and renormalization of
gravitational couplings}

Let us discuss the renormalization in more detail.
Vacuum polarization in an external gravitational field $g_{\mu\nu}$
results in the appearance of a non-trivial right-hand side in the
Einstein equations, the average value
of the stress energy tensor of a quantum field,
$\langle \hat{T}_{\mu\nu}\rangle$. Such equations
can be obtained as an extremum
of an effective action $\Gamma[g]$
under variation of the background metric $g_{\mu\nu}$.
The effective action has the following form
\begin{equation}\label{3.4}
\Gamma [g]= I [g] +W[g]~~,
\end{equation}
where $I[g]$ is the classical Einstein action or its
modifications
and $W[g]$ is a functional related to the contribution
of quantum fields. For instance, for the so called
(free) non-minimally coupled scalar
field $W=\frac 12 \ln \det(-\nabla^2+\xi R+m^2)$, where
$\xi$ is the constant of the coupling with the scalar curvature $R$.

Computations show that  $W[g]$
has ultraviolet divergences which
can be absorbed by
a renormalization of the couplings in the classical
action $I[g]$. To this end the latter is chosen in the
form
\begin{equation}\label{b}
I(G_B,\Lambda_B,c^i_B)=\int d^4x \sqrt{g}
\left[ -\frac{\Lambda_B}{8\pi G_B} -\frac{R}{16\pi G_B}
+c^1_B R^2 +c^2_B R_{\mu\nu}R^{\mu\nu} +c_B^3
R_{\alpha\beta\mu\nu}R^{\alpha\beta\mu\nu}
\right]~~~.
\end{equation}
Denote by
$W_{\tiny\mbox{div}}$ the UV-divergent part of the quantum
action $W$. Then the renormalized quantities are
defined as
\begin{equation}
I_{\tiny\mbox{ren}}\equiv I(G_{\tiny\mbox{ren}},\Lambda_{\tiny\mbox{ren}},
c^i_{\tiny\mbox{ren}})=I(G_B,\Lambda_B,c^i_B)+
W_{\tiny\mbox{div}}~~~,\hspace{0.5cm}
W_{\tiny\mbox{ren}}=W-W_{\tiny\mbox{div}}~~~.
\end{equation}
The key observation is that
$W_{\tiny\mbox{div}}$ has the same structure as (\ref{b}) and
hence $W_{\tiny\mbox{div}}$ can be absorbed   by
simple
redefinition of the coupling constants in
$I(G_B,\Lambda_B,c^i_B)$. In other words,
$I_{\tiny\mbox{ren}}$ is identical to the initial classical action
$I$ with the only change that the bare coefficients
$\Lambda_B$, $G_B$, and $c^i_B$ are replaced by their renormalized
versions $\Lambda_{\tiny\mbox{ren}}$, $G_{\tiny\mbox{ren}}$, and
$c^i_{\tiny\mbox{ren}}$.
The relation between bare and renormalized couplings depends
on the regularization.
For instance, in PV regularization
the renormalization of the Newton constant for the non-minimally
coupled scalar field is
\begin{equation}\label{x1}
{1\over G_{\tiny\mbox{ren}}}={1\over G_{B}}+{\lambda \over 2\pi}
\left(\frac 16-\xi\right)\mu^2~~~
 ,
\end{equation}
where $\lambda=\ln{729 \over 256}$ and $\mu$ is the PV cutoff.
According to the general prescription, the observable constants
are identified with the renormalized ones. Thus, the Bekenstein-Hawking
entropy is $S^{BH}=
S^{BH}(G_{\tiny\mbox{ren}})={\cal A}/(4G_{\tiny\mbox{ren}})$.
As follows from (\ref{3.3}) and (\ref{x1}) it can be written in the
following form
\begin{equation}\label{3.5}
S^{BH}(G_{\tiny\mbox{ren}})=
S^{BH}(G_{\tiny\mbox{B}})+S(\mu)-
Q_{\tiny\mbox{div}}~~~,
\end{equation}
where $Q_{\tiny\mbox{div}}=\xi \lambda \mu^2 {\cal A}/(2\pi)$.

Equation (\ref{3.5}) explicitly demonstrates
that the ``observable'' Bekenstein-Hawking entropy contains the
statistical-mechanical entropy $S(\mu)$ of
the black hole's quantum excitations
as an essential part.  It can be shown \cite{DLM:95},\cite{FS:96}--\cite{Furs:98}
that this result does not depend on the
regularization procedure, or on the black hole background and
holds for the entropies of different fields. In general,
equation (\ref{3.5}) is extended to include corrections to the
Bekenstein-Hawking entropy due to terms depending on curvatures
\cite{FS:96}--\cite{Solodukhin:95}.

Relation (\ref{3.5}) removes the two problems formulated in the end of
section 3.1. Indeed, if one has a gravity theory where computations are
based on the renormalization procedure the leading part
of the entanglement entropy of all quantum excitations  is
just a part of the observable Bekenstein-Hawking entropy, no matter
what kind of regularization is used
and how many field species exist in the Nature.

Although this fact indicates a strong connection
between the entanglement entropy and $S^{BH}$
it does not solve the problem of the Bekenstein-Hawking entropy.
Indeed, one part of the observable entropy $S^{BH}$
is the "bare entropy" $S^{BH}(G_{\tiny\mbox{B}})$ in (\ref{3.5}) which has no
statistical-mechanical meaning.
Another question concerns statistical meaning of quantity $Q_{\tiny\mbox{div}}$
which appears due to non-minimal couplings.

\subsection{Induced gravity}

As was pointed out in  \cite{Jaco:94}, \cite{FFZ:97}, \cite{FF:97},
the problem of the bare entropy in (\ref{3.5}) can be resolved
if Einstein gravity is entirely induced by quantum
effects.
The idea of induced gravity was formulated by Sakharov
\cite{Sakh:68}, \cite{Sakh:76} and then developed in different
works, see, e.g. \cite{Adler:82} and the review papers \cite{NoVa:91},
\cite{Viss}. Sakharov's idea is very simple and physical.
Its main assumption is that the dynamical equations for the
gravitational field $g_{\mu\nu}$  are determined by properties
of the physical vacuum which, like an ordinary medium, has a
microscopic structure.  The relevant example is a crystal
lattice. The metric $g_{\mu\nu}$
plays the same role as the strength tensor  
$\sigma_{ij}=\frac 12 (\xi_{i,j}+\xi_{j,i})$ which describes macroscopic
deformations of a crystal
(here $\xi_i=\xi_i({\bf x})$ is a vector of the displacement of the
lattice site at a point with the coordinates $\bf{x}$).
Gravitons in this picture are analogous to phonons and are collective
excitations of the microscopic degrees of freedom of the vacuum.
We call these degrees of freedom {\it constituents}. The constituents
are virtual particles of all possible fields present in Nature.
The energy stored in the deformation of the crystal has the form
\begin{equation}\label{3.6}
{\cal E}[\sigma]=\sum_{\bf x}\left(A\sigma_{ii}^2({\bf x})+
B\sigma^{ij}({\bf x})\sigma_{ij}({\bf x})\right)~,
\end{equation}
where the coefficients $A$ and $B$ are determined by the
microscopic structure
of the lattice. They are known as Young and Poisson constants.
The physical vacuum responds to
variations of the metric $g_{\mu\nu}$
in a similar way. Such quantum effects can be described with the help
of the effective gravitational action
$\Gamma[g]$,
\begin{equation}\label{3.7}
e^{i\Gamma[g]}=\int [D\Phi] \exp (iI[\Phi,g ]) ,
\end{equation}
where the integration runs over
all constituent fields (denoted by $\Phi$).
$I[\Phi,g ]$ is the classical action of $\Phi$ on a classical
background with the metric $g_{\mu\nu}$.
Sakharov's idea is based on the observation that the leading
contribution to
$\Gamma[g]$ is determined by the divergent part and
has the form of the classical Einstein action
\begin{equation}\label{3.8}
\Gamma[g]\simeq {1 \over 16\pi G_{\tiny{\mbox{ind}}}}\int\sqrt{g} d^4x
(R(g)-2\Lambda_{\tiny{\mbox{ind}}}).
\end{equation}
Here  and in what follows we consider four-dimensional gravity.
$G_{\tiny{\mbox{ind}}}=({\gamma \over l^2})^{-1}$  is an
{\it induced} gravitational coupling, $\gamma$ is a numerical coefficient
which depends on the specific set of constituents and $l$
is a cut-off parameter in the region of high energies.
$\Lambda_{\tiny{\mbox{ind}}}$ is an induced cosmological constant.
It follows from  (\ref{3.6}) and (\ref{3.8}) that
$\Gamma[g]$ is similar to the energy
${\cal E}[\sigma]$, while  $G_{\tiny{\mbox{ind}}}$ and
$\Lambda_{\tiny{\mbox{ind}}}$ appear in the same way as Young and
Poisson constants.

There are very interesting parallels between induced gravity
and condensed matter systems, such as superfluid 3He and
4He, see \cite{VoZe:03}.

As was pointed out by Weinberg \cite{Wein} an analog of induced
gravity can be also found in particle physics.
It is a theory of soft pions which can be used
in the limit when the masses of the $u$ and $d$ quarks are neglected.
In this limit there is a global chiral
$SU(2)\times SU(2)$ symmetry.
The gravitational action (\ref{3.8}) has the same meaning as the Lagrangian
of the chiral model while
the constituents are analogous to the quarks
that the pions are made of.

Let us note that in the interval of low energies, as in the case of
pions, one can develop a quantum theory
of gravitons by using (\ref{3.8}). This theory
can be used, for instance,
to calculate graviton scattering
amplitudes, see \cite{Burg:03} for discussion
of this topic.

The problem of the statistical interpretation of the Bekenstein-Hawking
entropy in induced gravity is resolved in the following way.
The microscopic degrees of freedom  responsible for
$S^{BH}$ are the constituents which live in the gravitational
field of  a black hole. These virtual particles have a non-trivial
quantum stress-energy tensor $\langle \hat{T}_{\mu\nu} \rangle$
which can be obtained by variation of the induced effective
action (\ref{3.7}). The background metric is a solution
of the equation
\begin{equation}\label{3.9}
\langle \hat{T}_{\mu\nu} \rangle=0~~.
\end{equation}
In the limit when the gravitational
radius
is much larger than the Planck length
$L_{Pl}=G^{1/2}_{\tiny{\mbox{ind}}}$
the effective action reduces to (\ref{3.8}) and equations (\ref{3.9})
reduce to the Einstein vacuum equations.
Because the black hole is a solution of these equations its entropy is
$S^{BH}=
{\cal A} /4 G_{\tiny{\mbox{ind}}}\sim {\cal A} /l^2$ and it has
the same order of magnitude as the entropy of the constituents
near the horizon computed with the use of the cutoff $l$.

\subsection{Induced gravity models}

The above explanation of the Bekenstein-Hawking entropy is rather
schematic because it implies the existence of a cutoff mechanism in
the region of high energies which we do not know.

To verify whether induced gravity can really explain the black hole
entropy one needs additional assumptions and concrete models.
This step was carried out in \cite{FFZ:97}, \cite{FF:97} where the additional
condition was that the theory of
constituents was free from leading ultraviolet divergences.
This requirement
enables one to construct models where $G_{\tiny{\mbox{ind}}}$
is a computable quantity.

Induced gravity models having this property
may possess different types of
constituent fields. We consider the simplest possibility.
The model consists of $N_s$ scalar constituents $\phi_s$
with masses $m_s$, some
of the constituents being non-minimally coupled to the background
curvature with corresponding couplings $\xi_s$, and $N_d$ Dirac fields
$\psi_d$ with masses $m_d$. The corresponding actions
in (\ref{3.7}) are
\begin{equation}\label{3.10}
I[\phi_s, g]=-\frac 12 \int d^4x \sqrt{-g} \left[(\nabla \phi_s)^2+
\xi_s R\phi_s^2+m_s^2\phi_s^2\right]~,
\end{equation}
\begin{equation}\label{3.11}
I[\psi_d, g]=\int d^4x \sqrt{-g}~ \bar{\psi}_d(i\gamma^\mu
\nabla_\mu+m_d)\psi_d~.
\end{equation}
Let us impose the following constraints on parameters
of the constituents:
\begin{equation}\label{3.12}
p(0)=p(1)=p(2)=p'(2)=0~,
\end{equation}
\begin{equation}\label{3.13}
q(0)=q(1)=0~,
\end{equation}
where
\begin{equation}\label{3.14}
p(z)=\sum_s m_s^{2z}-4\sum_d m_d^{2z}~,\hskip 1cm
q(z)=\sum_k c_k m_k^{2z}~,
\end{equation}
$k=s,d$, and $c_d=2$, $c_s=1-6\xi_s$ for spinor and scalar
constituents, respectively.
The constraints (\ref{3.12}) serve to eliminate
the induced cosmological constant while (\ref{3.13})
enable one to get rid of the ultraviolet divergences
in the induced Newton constant $G_{\tiny{\mbox{ind}}}$.
It is  the second set of
conditions that will be important for our analysis of black hole
entropy.
Given (\ref{3.13}) $G_{\tiny{\mbox{ind}}}$
is defined by the formula
\begin{equation}\label{3.15}
{1 \over G_{\tiny{\mbox{ind}}}}={1 \over 12\pi}q'(1)=
{1 \over 12\pi} \sum_k c_k m_k^2\ln m_k^2~.
\end{equation}
Because $G_{\tiny{\mbox{ind}}}$ is explicitly known one can prove that
\begin{equation}\label{3.16}
S^{BH}={{\cal A} \over 4G_{\tiny{\mbox{ind}}}}=S-Q~.
\end{equation}
Here $S$ is a statistical-mechanical entropy of the constituents
thermally distributed at the Hawking
temperature in the vicinity of the horizon .
The quantity
$Q$ is a quantum average of the operator
\begin{equation}\label{3.17}
\hat{Q}=2\pi \sum_s \xi_s \int_{\Sigma} d^2\sigma \hat{\phi}_s^2 ~
\end{equation}
where the integration goes over the horizon surface $\Sigma$.

The reason why a quantity like $Q$ appears in the entropy formula
is the following. The constraints (\ref{3.13}) cannot be satisfied
without introduction of non-minimal couplings $\xi_s R\phi_s^2$
in the scalar sector, see (\ref{3.10}). $G_{\tiny{\mbox{ind}}}$ and
$S^{BH}$ depend on the non-minimal
coupling constants $\xi_s$ while the thermal entropy
$S$ does not. This disagreement in the behavior of the two entropies
appeared already in the renormalization equation (\ref{3.5}).
What happens in (\ref{3.16}) is that the divergence in $S$ is compensated
by the divergence in $Q$.

Formula (\ref{3.17}) is rather universal: it is valid for different models
including those with vector constituents \cite{FF:98v}
as well as for different kinds of black holes,
rotating \cite{FF:99kerr} and charged
\cite{FF:99ch}, in different space-time
dimensions.

\bigskip

What can one learn from these results?

i) The induced gravity models give a
physical picture of the microscopic
degrees of freedom of a black hole responsible for its entropy.
These degrees of freedom are the constituents propagating near the
black hole horizon. The source of the Bekenstein-Hawking entropy
is the entanglement or thermal entropy of the constituents
in the given black hole background.

ii) Induced gravity is not a fundamental theory but has the key
properties which an ultimate theory of quantum gravity must possess.
These properties are: the generation of the equations of the gravitational
field by quantum effects and the absence of
the leading ultraviolet divergences.  As was pointed out in
\cite{HaMaSt:01},
from the point of view of the open string theory
black hole entropy can be considered as a loop effect, in full
analogy with its origin in induced gravity.

\bigskip

To summarize,  induced gravity is an example where one can
study the mechanism of generation of the Bekenstein-Hawking entropy
by using very general properties of a hypothetical fundamental
theory.

\bigskip

The question which is not completely resolved
in induced gravity models is the statistical
meaning of quantity
$Q$ in (\ref{3.16}). Since this term is present it is not quite clear how to
represent $S^{BH}$  in the form $-\mbox{Tr}\hat{\rho}\ln \hat{\rho}$
(see, however, \cite{Furs:00b}).

The physical reason of subtracting $Q$ in (\ref{3.16}), as was explained
in \cite{FF:97}, is related to two inequivalent definitions
of the energy in the black hole exterior. One definition, $H$,
is the canonical energy or the  Hamiltonian. The other definition,
$E$, is the energy expressed in terms of the stress-energy tensor
$T_{\mu\nu}$ which is obtained by variation of the action over the
metric tensor.  The two energies correspond to different
properties of a black hole. $H$ corresponds to evolution
of the system along the Killing time and for this reason
the operator $H$ in quantum theory is used for constructing the density
matrix which yields the entropy $S$ in (\ref{3.16}).
On the other hand, $E$ is related to thermodynamical properties of
a black hole. If the black hole mass measured at infinity is
fixed the change of the entropy $S^{BH}$ caused by the change of
the energy $E$ of fields in the black hole exterior is
\begin{equation}\label{3.18}
\delta S^{BH}=-T_H \delta E~,
\end{equation}
where $T_H$ is the Hawking temperature of the black hole.
The reason why $E$ and $H$ are not equivalent is in the existence of
the horizon. The two quantities
being integrals of metrical and canonical stress tensors
differ by a total derivative. This difference
results in a surface term (a Noether charge) on the bifurcation
surface of the
horizon. This surface term does not vanish because the horizon is
not a real boundary and the only requirement for fields in this
region is regularity. One can show \cite{Furs:99} that the boundary
term is the $Q$ appearing in (\ref{3.16}). More
precisely,
\begin{equation}\label{3.19}
E=H-T_H Q~.
\end{equation}
According to (\ref{3.18}) the black hole entropy is related to the
distribution over the energies $E$ of the induced gravity constituents.
Hence, the subtraction of $Q$ in (\ref{3.16}) accounts for the difference
between $E$ and $H$ in (\ref{3.19}).

It should be noted, however, that an explicit calculation of the
black hole degeneracy for a given mass $M$ which is connected with
the distribution of the constituent field states over the
energies $E$ is a problem. Two suggestions how it can be done are
discussed in \cite{FF:97} and \cite{Furs:00b}. The difficulty is that in
quantum theory a non-zero value of $Q$ in (\ref{3.16}) is ensured by
modes which, from the point of view of a static observer, have
vanishing frequencies, the so-called soft modes.

\section{CFT and induced gravity}
\setcounter{equation}0
\subsection{Dimensional reduction}

We pointed out in section 2.6 that so far there are no examples
showing that the near-horizon
conformal symmetry can be realized in quantum gravity theory.
Before such examples are known one can investigate this
question in some simple models. This is another
case where
induced gravity can be quite helpful in developing ones intuition.
In this section we follow the work \cite{FFZ:03}.

Let us note that in the considered models the induced gravity
constituents are massive fields whose masses have to be comparable
to the Planck mass to ensure that the induced Newton constant
(\ref{3.15}) has the correct value. The contribution to
$G_{\tiny{\mbox{ind}}}$ from  the fields
observable at low-energies (fields of the Standard Model)
can be neglected. How can the presence of massive constituents
be reconciled with conformal symmetry? The idea is simple:
since the local temperature of quanta near the horizon is large
the fields living within certain distance to the horizon
are effectively massless
and scale invariant. The role of the masses is to introduce a scale
(a correlation length) into the CFT theory.

The curvature effects near the horizon are not important
and one can use approximation (\ref{2.24}) where the metric
on the horizon itself is replaced by the flat metric
\begin{equation}\label{4.1}
d\sigma^2=dy_1^2+dy_2^2
~~.
\end{equation}
The conformal transformations change only the metric $d\gamma^2$ in
two-dimensional plane ${\cal G}$ orthogonal to the horizon surface.
We will
write this metric as
\begin{equation}\label{4.2}
d\gamma^2=\gamma_{\alpha\beta}dx^\alpha dx^\beta~,
\end{equation}
where $\alpha,\beta=0,1$ and for a moment let
$\gamma_{\alpha\beta}$ be arbitrary.

In this setting the dynamics of the constituents is essentially
two-dimensional. This can be easily seen if we use the
Fourier decomposition in $y$-plane and define
\begin{equation}\label{4.3}
\Phi_{{\bf p}}(x)={1 \over 2\pi a}\int d^2y~
e^{-i\bf{py}}\Phi(x,y)~,
\end{equation}
where $\bf p$ is a momentum along the horizon,
${\bf p}{\bf y}=p_i y^i$.
To avoid volume divergences related to the infinite size of the
horizon we  assume that the range of coordinates $y^i$
is restricted,
$-a/2\leq y^i \leq a/2$. This means that the horizon area
${\cal A}$ is finite and equal to $a^2$.

Thus, each 4D field $\Phi(x,y)$  corresponds
to a tower of 2D fields $\Phi_{{\bf p}}(x)$ which live on
${\cal G}$. If $\Phi(x,y)$ has the mass $m$
then the mass of $\Phi_{{\bf p}}(x)$ depends on the
transverse momentum ${\bf p}$,
\begin{equation}\label{4.4}
m({\bf p})=\sqrt{m^2+{\bf p}^2}~.
\end{equation}
It should be noted that if the induced gravity constraints
(\ref{3.12}), (\ref{3.13}) are satisfied for the set of masses
$m_s$, $m_d$, they are satisfied for the masses $m_s({\bf p})$,
$m_d({\bf p})$ as well. This means that a 2D gravity theory induced in
each 2D sector at a given transverse momentum ${\bf p}$ is free
from UV divergences.
The effective action $\Gamma[g]$ of the 4D induced gravity
is the sum of the actions  $\Gamma_2[\gamma,p]$ of 2D gravities
\begin{equation}\label{4.5}
\Gamma[g]=\sum_{\bf p} \Gamma_2[\gamma, p]
\simeq {a^2 \over 4\pi}\int^\infty_\sigma \Gamma_2[\gamma,p] dp^2 ~.
\end{equation}
Here $p=|{\bf p}|$.
It is assumed in  (\ref{4.5}) that the parameter $a$ is large, so
the sum over $\bf p$  replaced by the integral over $p$.
The coefficient $a^2 /(4\pi)$ is related to
the number of modes with the momentum square $p^2$.

The two-dimensional action can be easily calculated,
\begin{equation}\label{4.6}
\Gamma_2[\gamma,p]\simeq{1 \over 4G_2(p)} \int\sqrt{-\gamma} d^2x
~\left({\cal R}+2\lambda_2(p)\right)~.
\end{equation}
Here ${\cal R}$ is the curvature of  ${\cal G}$, and
\begin{eqnarray}\label{4.7}
{1 \over G_2(p)}&=&-{1 \over 12\pi}
\sum_k c_k\ln m_k^2({\bf p}) ~,\\
\label{4.8a}
{\lambda_2(p) \over G_2(p)}&=&{1 \over 4\pi}
\left[ \sum_s m_s^2({\bf p})\ln m_s^2({\bf p}) -4 \sum_d m_d^2({\bf p})
\ln m_d^2({\bf p})\right]~.
\end{eqnarray}

The four-dimensional Newton constant $G_{\tiny{\mbox{ind}}}$ can
be found by summation over momenta in (\ref{4.5}) if one takes into
account that $a^2=\int dy_1 dy_2$.
It gives $\Gamma[g]$
in the form (\ref{3.8}), where $R[g]={\cal R}[\gamma]$ and
\begin{equation}\label{4.8}
{1 \over G_{\tiny{\mbox{ind}}}}=\lim_{p \rightarrow 0}{1 \over G(p)}~,\hskip 2cm
{1 \over G(p)}=\int_{p^2}^\infty {d{\tilde p}^2 \over G_2({\tilde p})}~,
\end{equation}
which coincides with (\ref{3.15}).

Let us make an additional assumption:
we treat the two-dimensional field models at
any momentum $p$ not just Fourier components but as {\it physical
theories} in a sense that each of these theories yields
a 2D induced gravity with strictly
positive gravitational couplings $G_2(p)$.
In this case
the Bekenstein-Hawking entropy of a black hole in such a 2D
gravity is positive. Examples of induced gravity models
with this property are presented in \cite{FFZ:03}.

\subsection{Representation of the near-horizon CFT}

The 2D constituent fields
create a thermal atmosphere around  a 2D black hole,
see the discussion in section 3.1. This entropy can be easily computed
if we neglect
for a moment the masses of the fields and the non-minimal couplings.
Near the horizon  the 2D metric $d\gamma^2$  is the 2D Rindler
metric (\ref{2.25}). To avoid divergences near the horizon
we introduce a cutoff $\epsilon$ by imposing a restriction
in  (\ref{2.25}) $\epsilon \leq \rho \leq R$.
The upper cutoff $R$ is  needed to eliminate an infrared
divergence at spatial infinity. Since the theory is scale invariant
one can rescale the metric (\ref{2.25}) to the form
\begin{equation}\label{4.9}
d\tilde{\gamma}^2=-dt^2+dx^2
~~,~~x={1 \over \kappa}\ln \rho~~.
\end{equation}
Therefore, the theory we are dealing with is a massless 2D field
on an interval of the length $b=\kappa^{-1}\ln (R/\epsilon)$.
To find its entropy one can use the result (\ref{1.7})
\begin{equation}\label{4.10}
S=S(T_H,b)\simeq {\pi \over 3}bT_H=\frac 16 \ln {R \over \epsilon}~~,
\end{equation}
where we took into account that the temperature has to be
identified with the temperature of the Hawking radiation.

How do masses and non-minimal couplings change this result?
As was shown in \cite{FFZ:03}, one can formulate the following rules:

i) Near the horizon each induced gravity constituent with the momentum
$p$ and mass $m_k$
corresponds to a 2D conformal theory
on an interval $b_k=\kappa^{-1}\ln (R_k(p)/\epsilon)$ where the external
radius is determined as
$R_k(p)=m_k({\bf p})^{-1}$, $p=|{\bf p}|$;

ii) Each of these CFT's is characterized by a central charge $c_k$;
charges of spinor constituents are
$c_d=2$, while charges of scalar fields are $c_s=1-6\xi_s$ and
depend on the non-minimal couplings.

\bigskip

The first rule  follows from the fact that the two-point correlator
of field operators is exponentially small when fields are separated
by a distance larger than their correlation length $m_k({\bf p})^{-1}$.
The second rule can be inferred from the transformation properties
of the components of the renormalized stress-energy tensor
of 2D fields.
For example, for a scalar field with the non-minimal coupling $\xi$
the $uu$ component of the stress-energy tensor
\begin{equation}\label{4.11a}
T_{uu}=\langle -(\partial_u \hat{\phi})^2+
2\xi((\partial_u \hat{\phi})^2+\hat{\phi} \partial_u^2 \hat{\phi}) \rangle~
\end{equation}
transforms as
\begin{equation}\label{4.11}
\delta T_{uu}= \varepsilon \partial_u T_{uu}+
2\partial_u\varepsilon T_{uu} +{1-6\xi \over 24\pi}
\partial_u^3\varepsilon +O(\varepsilon^2)
\end{equation}
under an infinitesimal change $\delta u=\varepsilon(u)$
of the light cone coordinate $u$ (the light-cone coordinates
are introduced in (\ref{2.25})).
Eq. (\ref{4.11}) has the same form as transformation
(\ref{2.8a}) in a CFT theory with the central charge $c=1-6\xi$.

The induced gravity constraints (\ref{3.13}) which
eliminate the divergences in the induced Newton constant
$G_{\tiny{\mbox{ind}}}$
can be represented in the form
\begin{equation}\label{4.12}
\sum_k c_k=0~,~~~\sum_k c_km_k^2=0
\end{equation}
The sum $C=\sum_k c_k$ can be interpreted as a total central charge
of the constituents.
This charge is zero because at each momentum $p$
the 2D theory is free from ultraviolet
divergences.

Note that (\ref{4.12}) requires that some
central charges $c_s$ are negative.
Typically CFT's with negative central charges correspond to ghosts.
The ghosts appear in gauge theories when the
Hilbert space is enlarged during
quantization. Ghosts give negative contributions to the
entropy to
compensate for the
contribution of the extra degrees of freedom in the
enlarged Hilbert space. However, if the system  is unitary its
total entropy is always positive.
As for ghost fields, the entropy associated with the 2D constituents
with negative $c_k$ is negative, and as in gauge theories
the total entropy in each 2D induced gravity sector
is positive because of
the requirement $G_2(p)>0$.

\bigskip

Now one can construct a concrete representation of the algebra of conformal
transformations in the $\rho-t$ plane in terms of the operators
acting in a Fock space of the CFT's. This can be
used to count the degeneracy of states corresponding to certain
energy levels as is done in the near-horizon approach discussed
in section 2.6. Instead of doing this we give a simpler
derivation based on equation (\ref{4.10}). According
to the formulated rules, each 2D constituent gives the following
contribution
\begin{equation}\label{4.13}
s(c_k,b_k(p))={c_k \over 6} \ln {R_k(p) \over \epsilon}=
-{c_k \over 6} \ln \epsilon m_k(p) ~~
\end{equation}
to the total entropy.
The entropy of all constituents in 2D
induced gravity at some momentum $p$ is
\begin{equation}\label{4.14}
s(p)=\sum_k s(c_k,b_k(p))=
{1 \over 6} \sum_k c_k \ln R_k(p)={\pi \over G_2(p)}~,
\end{equation}
where $G_2(p)$ is the 2D induced Newton constant defined
in (\ref{4.7}). The dependence on cutoff $\epsilon$ disappears
because of  (\ref{4.12}).
As was pointed out above, the  partial entropy $s(p)>0$ because
$G_2(p)>0$; $s(p)$ is just the entropy of a black hole
in the corresponding 2D induced gravity theory.
The entropy in the 4D theory is
\begin{equation}\label{4.15}
S_{\tiny{\mbox{tot}}}={a^2 \over 4\pi} \int_0^\infty s(p) dp^2=
{{\cal A} \over 4G_{\tiny{\mbox{ind}}}}~.
\end{equation}
It coincides with the Bekenstein-Hawking entropy (\ref{3.15})
of a four-dimensional
black hole with the horizon area ${\cal A}=a^2$.
The last equality in (\ref{4.14})
follows from  relation (\ref{4.8}) between 4D and 2D couplings.

\bigskip

Several remarks about this result are in order.

i) The given analysis shows that the method based on the near-horizon CFT
does reproduce the Bekenstein-Hawking entropy
in the induced gravity theory
and it has there a concrete realization.

ii)  It shows that the dimensional parameter which defines the "size"
$b$  of the near-horizon CFT may have a dynamical origin and is
related to physics at Planckian scales.

iii) The near-horizon CFT's are effective theories because they are
obtained as a result of dimensional reduction. The
definition of 2D fields depends on the horizon radius, see (\ref{4.3}).
Thus, black holes with different horizon areas are described
by different CFT's. This  property is similar to what one has
in the approach \cite{Carl:99a}, \cite{Solo:99}.
The  question of whether black hole evaporation may result in the information loss
should be addressed in the original theory of 4D constituents.

\bigskip

Apart from these similarities the near-horizon CFT
in induced gravity has several features which do not appear
in the approach discussed in section 2.6.

i) The total central charge in this theory vanishes, see (\ref{4.12}).
This property is related to cancellation of the leading
ultraviolet divergences.

ii) Because the masses of constituens are different, there is a set of
correlation lengths $m_k^{-1}({\bf p})$. Thus, such a theory may
possess several diffrent scales.

iii) What is important for understanding of the black hole entropy
is not only the conformal symmetry itself  but also the way it is
broken.

iv)  Interpretation of induced gravity  in terms of a near
horizon CFT requires further restrictions on the parameters
of constituents to ensure positivity of 2D gravitational coupling
$G_2(p)$ at each transverse momentum.

v) Each 2D induced gravity sector contains negative central charges.
As for ghost fields, the entropy associated with the corresponding
degrees of freedom has to be subtracted from the total entropy.
This property requires further analysis.

\bigskip

Finally, it should be noted that the
computations of $S^{BH}$ we discussed in this section
can be done not only in four-dimensional space-times, see \cite{FFZ:03}.
It would be very interesting to investigate other possibilities
of realizing the near horizon CFT in induced gravity.

\section{Summary}

We discussed several examples that strongly support the idea
that a microscopic origin of the Bekenstein-Hawking entropy
of black holes can be understood by using a few general properties
of a fundamental quantum gravity theory.  These properties
may be gleaned entirely  from  low-energy physics.

One of the possibilities is that finding a proper place for
a group of 2D conformal symmetries will make it possible to
control completely the density of states in quantum gravity.
The other possibility is that the entropy of a black hole
can be considered as a measure
of the information loss inside the horizon provided that the
gravity is entirely induced by quantum effects and the underlying
theory is ultraviolet finite. These two points of view may
complement each other.

It is fair to say that although these possibilities are very promising,
both approaches have unresolved problems.
It is a matter
for future research to see whether the discussed
problems are technical or whether
they are more fundamental and, hence, require something which we
cannot know about quantum gravity at low energies.

\noindent
\section*{Acknowledgments}

The author is very grateful to V. Frolov, W. Israel and A. Zelnikov
for reading the manuscript and helpful comments.

\end{document}